\documentclass[sigconf]{acmart}

\usepackage{graphicx}

\usepackage{multirow}
\usepackage{balance}
\balance

\begin{document}
\title[Quality Assurance of Generative Dialog Models]{Quality Assurance of Generative Dialog Models in an Evolving Conversational Agent Used for Swedish Language Practice}

\author{Markus Borg}
\orcid{XXX}
\affiliation{%
  \institution{RISE Research Institutes of Sweden}
  \city{Lund}
  \country{Sweden}
}
\email{markus.borg@ri.se}

\author{Johan Bengtsson}
\orcid{XXX}
\affiliation{%
  \institution{Lund University}
  \city{Lund}
  \country{Sweden}
}
\email{jo6064be-s@student.lu.se}

\author{Harald Österling}
\orcid{XXX}
\affiliation{%
  \institution{Lund University}
  \city{Lund}
  \country{Sweden}
}
\email{ha8034os-s@student.lu.se}

\author{Alexander Hagelborn}
\orcid{XXX}
\affiliation{%
  \institution{NordAxon AB}
  \city{Malm\"o}
  \country{Sweden}
}
\email{alexander.hagelborn@nordaxon.com}

\author{Isabella Gagner}
\orcid{XXX}
\affiliation{%
  \institution{NordAxon AB}
  \city{Malm\"o}
  \country{Sweden}
}
\email{isabella.gagner@nordaxon.com}

\author{Piotr Tomaszewski}
\orcid{XXX}
\affiliation{%
  \institution{RISE Research Institutes of Sweden}
  \city{Lund}
  \country{Sweden}
}
\email{piotr.tomaszewski@ri.se}

\renewcommand{\shortauthors}{Borg \textit{et al.}}

\begin{abstract}
Due to the migration megatrend, efficient and effective second-language acquisition is vital. One proposed solution involves AI-enabled conversational agents for person-centered interactive language practice. We present results from ongoing action research targeting quality assurance of proprietary generative dialog models trained for virtual job interviews. The action team elicited a set of 38 requirements for which we designed corresponding automated test cases for 15 of particular interest to the evolving solution. Our results show that six of the test case designs can detect meaningful differences between candidate models. While quality assurance of natural language processing applications is complex, we provide initial steps toward an automated framework for machine learning model selection in the context of an evolving conversational agent. Future work will focus on model selection in an MLOps setting. 
\end{abstract}

\copyrightyear{2022} 
\acmYear{2022} 
\acmConference[CAIN '22]{International Conference on AI Engineering}{May 22--23, 2022}{Pittsburgh, USA}

%
%



\maketitle

\section{Introduction}
Migration is a megatrend of our time. In the ongoing globalization, the number of international migrants has reached unprecedented heights. International migration is a complex phenomenon that includes both \textit{voluntary migration}~\cite{ottonelli2013migration}, e.g., to enhance career opportunities, and \textit{forced migration}, e.g., caused by conflicts~\cite{yazgan2015syrian} and climate changes~\cite{abel2019climate}. While the Covid-19 pandemic has temporarily slowed migration flows, the UN's position is that ``migration is part of today's world’’~\cite{un_2020}. The UN expects the numbers to rise again and posits that well-managed migration can contribute to inclusive and sustainable development in countries of both origin and destination. 

Efficient and effective second-language acquisition is a vital component in sustainable international migration. The body of research literature points out language proficiency as a key to successful integration~\cite{blake2017impact,meer2019english}, not the least to support rapid entry to the labor market. International migrants are a highly diverse set of people with substantially different capacities, needs, and opportunities for learning~\cite{morrice2021you}. The history of second-language acquisition has progressed from a focus on linguistics, e.g., inspired by Chomsky's ideas about a universal grammar~\cite{chomsky2014aspects}, to a sociocognitive process, i.e., learning by combining cognitive involvement and social interaction~\cite{larsen2018looking}. The current school of thought stresses the importance of person-centered interactions. While such interactions undoubtedly offer effective learning environments, it remains unclear how they could scale to meet the needs of the migration megatrend. 

Digital solutions for scalable second-language acquisition are increasingly popular. As there is a high demand for language learning tools, several applications have entered the consumer market in the last decade\footnote{https://blog.vsoftconsulting.com/blog/7-of-the-best-language-learning-chatbot-apps}. Duolingo, one of the most popular applications, has hundreds of millions of global users that are incentivized to keep learning through gamification~\cite{shortt2021gamification}. While Duolingo relies primarily on decontextualized grammar-translational exercises and audiolingual drills, there are indications that already such a simplistic digital interaction between a learner and a digital solutions can support second-language acquisition~\cite{loewen2019mobile}. Chatbots are another type of language learning tools~\cite{pham2018chatbot}, available in Duolingo and other services such as Andy, Eggbun, and Babbel. While chatbots are closer to the person-related interactions recommended for second-language acquisition, we focus this study on a more advanced digital solution, i.e., conversational agents (ConvAg).

State-of-the-art ConvAgs rely on generative dialog models (GDM). A GDM is a type of machine learning (ML) model that aims to produce human-like replies to user input. While many simple chatbots rely on pattern matching and string processing~\cite{hussain2019survey}, the GDMs we consider in this work are open-domain~\cite{roller2020recipes}, i.e., they shall be able to maintain a dialog on arbitrary topics. This is clearly a significantly more challenging implementation task and, as we stress in this paper, a substantially tougher challenge for any quality assurance (QA) effort. The AI engineering community needs to develop approaches for QA of GDMs, as failures can have severe repercussions. We remind the reader of Microsoft’s experimental ConvAg Tay, removed from Twitter less than 24 hours after being deployed due to posting racist, sexist, and anti-Semitic tweets~\cite{wolf2017we}.

In this paper, we argue that QA of ConvAgs' GDMs must be tackled from two directions. First, the development organization must capture the often vague expectations on how a ConvAg shall perform in its operational context. Second, test engineers must develop automated test cases that verify that the expectations are met. To exacerbate the QA challenge, typical ConvAg development involves frequent retraining of the underlying GDM. ML development is highly exploratory and known to suffer from the CACE principle~\cite{sculley_hidden_2015}, i.e., ``Changing Anything Changes Everything.'' Often several models are retrained in parallel, and then the ML engineers select one for deployment -- so called \textit{model selection}. While there might be indications that one of the models is better, how can you know for sure? We need a reliable approach for benchmarking different models~\cite{hasselbring2021benchmarking}, i.e., test automation that helps us detect if any GDMs digress from acceptable behavior.

The AI engineering community needs more research on QA of ML models trained for natural language processing (NLP). From the ML testing perspective, previous secondary studies show that NLP is underrepresented in the body of literature~\cite{riccio2020testing,zhang_machine_2020}. The need for NLP testing was also highlighted by Yoo in his keynote address at the DeepTest~2020 Workshop as he called for additional research~\cite{yoo2020searching}. We acknowledge that QA of GDMs forces us to see the oracle problem~\cite{barr2014oracle} in a new light. Also, compared to the comparatively well-studied topic of testing of ML models for computer vision,  the discrete nature of natural language~\cite{ferrone2020symbolic} poses new challenges, as a simple negation in a sentence inverts the meaning.

We present results from ongoing action research in the context of QA of GDMs for , an evolving ConvAg designed for Swedish language practice. Emely evolves as part of an innovation project in the city of Helsingborg under development of a team from the AI consultancy company NordAxon. During the summer of 2021, researchers from RISE and Lund University joined to constitute a QA action team~\cite{staron2020action}. 
Together, we set out to explore three research questions in the context of ConvAg QA:

\begin{itemize}
\item[RQ1] What are the requirements on Emely's GDM?
\item[RQ2] How can test cases be designed to verify that the GDM satisfies the requirements?
\item[RQ3] How can the test cases be implemented to allow test automation throughout the evolution of Emely?
\end{itemize}

Our pioneering paper responds to calls for additional research on ML testing for NLP. To the best of our knowledge, this is the first study to explore QA of a ConvAg development project in the field. In this paper, we present intermediate results and lessons learned from the first phase of our study. As Emely is a proprietary product, we complement our discussion with Facebook's open-source ConvAg Blenderbot~\cite{roller2020recipes}. The test infrastructure we use for GDM testing is available on GitHub~\cite{repo}, but only the Blenderbot part is available for external replication.

Note that the requirements reported in this paper are specified without strict quality targets for two reasons. First, the exact threshold values for Emely are business-critical and cannot be shared publicly. Second, Emely development follows the agile practice of using test cases as requirements~\cite{bjarnason2016multi}. This practice supports integration of requirements engineering in the agile development context, i.e., exact threshold values can easily be tuned by developers updating the automated test cases. As there are no guidelines to base the quality targets on, we found an experimental manner to be the best way forward.

The paper is organized as follows. First, Sec.~\ref{sec:bg} presents background and related work followed by a technical overview of Emely in Sec.~\ref{sec:case}. Second, Sec.~\ref{sec:method} describes our approach to action research and the results are discussed in Sec.~\ref{sec:res}. Third, Sec.~\ref{sec:threats} and Sec.~\ref{sec:lessons} elaborate on threats to validity and lessons learned, respectively, before we conclude the paper in Sec.~\ref{sec:conc}.

\section{Background and Related Work} \label{sec:bg}
This section introduces AI quality for ConvAgs. Furthermore, we present related work on NLP testing and GDM evaluation.

\subsection{AI Quality and ConvAgs} \label{sec:bg-q}
Expressing quality expectations on a ConvAg is far from trivial. First, ``quality'' in general is a notoriously difficult aspect to put the finger on~\cite{walkinshaw_software_2017}. Quality is not a single concept, but rather a multi-dimensional patchwork of different system aspects that influence the users' experience. Moreover, quality is inevitably subjective and largely lies in the eye of the beholder. For ConvAg quality, the user experience will vary depending on both the human conversationalist and the context, i.e., each dialog instance.

Second, ConvAgs are created to mimic a human conversationalist. This is indeed a grand task, and an ideal solution would even pass the imitation game of the famous Turing test~\cite{saygin2000turing}, i.e., the ConvAg would exhibit intelligent behaviour indistinguishable from that of a human. Assessing intelligent behavior is non-trivial, thus Turing proposed the role of a human interrogator in the original paper's envisioned ``imitation game.'' Pinpointing what is needed to pass the test, i.e., specifying requirements on what the interrogator should check, is a formidable challenge. Finally, as also humans are known to fail Turing tests~\cite{warwick2015human}, we humbly accept that ConvAg QA is a daunting task that cannot be solved in a single study.

In this paper, we use the following definitions from related work~\cite{borg2021aiq}: ``MLWare is a subset of AI that, fueled by data, realizes functionality through supervised
and/or unsupervised learning`` and ``AI Quality is the capability of MLware to satisfy stated and implied needs under specified conditions while the underlying data satisfy the requirements specific to the application and its context.'' A fundamental aspect of the latter definition is that quality is about both stated and implied needs. As part of our study, we conduct requirements engineering (RE) to transfer an initial set of implied needs on ConvAgs to specified requirements that lay the foundation for subsequent test case design and implementation. 

\subsection{NLP Testing and GDM Evaluation}
The software testing community has published relatively few papers on ML testing for NLP applications. Compared to ML testing for image classification and recognition, especially for automotive perception systems~\cite{borg_safely_2019}, few studies specifically target NLP~\cite{riccio2020testing,zhang2020machine}. Exceptions include a study by Kim and Yoo that explores how surprise adequacy, a test adequacy metric developed for neural network input~\cite{kim2019guiding}, generalizes from image input to the NLP context~\cite{kim2021multimodal}. The most similar work to ours is the dialog testing tool DialTest by Liu \textit{et al.}~\cite{liu2021dialtest}, which was not published when we initiated our action research study. 

In the NLP research community, several papers propose metrics to support evaluation of language generation. Deng \textit{et al.} presented a framework for natural language generation that covers language compression, transduction, and creation~\cite{deng2021compression}. The latter includes the dialog generation that constitutes a critical feature in ConvAgs. The authors' primary contribution is a set of metrics that corresponds to information alignment between natural language input, output, and the overall context. Based on an analysis of an established human annotation dataset, they conclude that the metrics correlate with human judgments. Yeh \textit{et al.} recently conducted a study comparing a set of 23 metrics for dialog generation on 10 datasets~\cite{yeh2021comprehensive}. They conclude that USR, GRADE, DEB, and USL-H perform the best for evaluating response generation.

Several researchers argue that a combination of metrics are needed to reliably evaluate generative language models. 
In the context of machine translation, Yuan \textit{et al.} propose BARTScore~\cite{yuan2021bartscore}, including evaluation based on seven quality categories: 1) Informativeness, 2) Relevance, 3) Fluency, 4) Coherence, 5) Factuality, 6) Semantic Coverage, and 7) Adequacy. These categories partly overlap the categories we found relevant for Emely, but we instead propose a customized list for GDMs.



\section{Case description: Emely} \label{sec:case}
\begin{figure}
    \centering
    \includegraphics[width=0.45\textwidth]{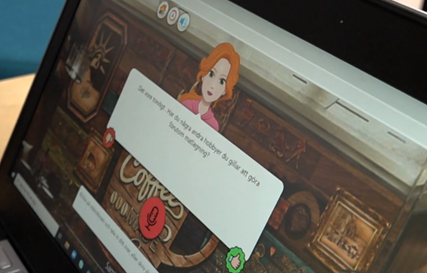}
    \caption{Emely's visual appearance.}
    \label{fig:web}
\end{figure}

Emely is a ConvAg designed to help newcomers in Sweden practice Swedish, i.e., to accelerate second-language acquisition. The long-term goal is to provide a scalable approach to Swedish language practice that will help migrants enter the labor market faster~\cite{blake2017impact,meer2019english}. Emely lets users practice job interview sessions in a protected setting, i.e., interactive person-centered dialogs as recommended for effective second-language acquisition~\cite{larsen2018looking}. Fig.~\ref{fig:web} provides an impression of Emely's visual appearance.

\begin{figure*}
    \centering
    \includegraphics[width=1\textwidth]{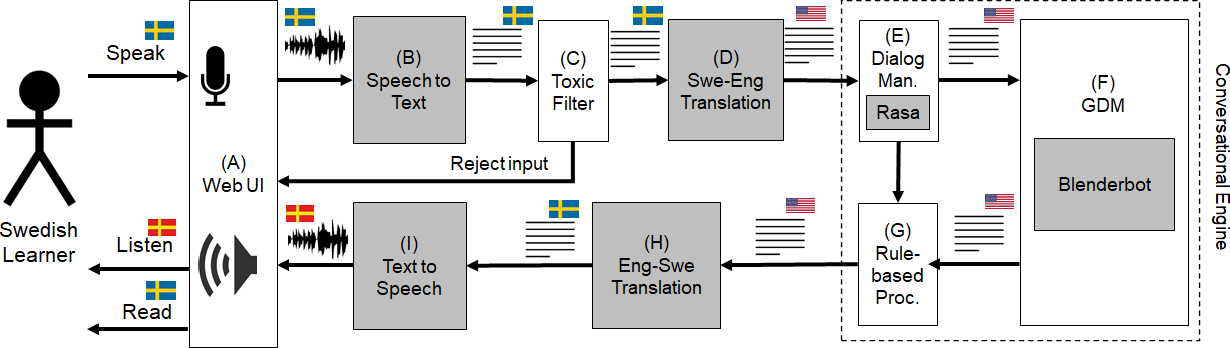}
    \caption{High-level Emely architecture. Gray boxes indicate third-party components.}
    \label{fig:architecture}
\end{figure*}

Fig.~\ref{fig:architecture} shows the high-level architecture of Emely. Note that Emely is a proprietary solution, thus we restrict the architecture description to a high level and quantitative information is reported in vague terms. Third-party components are depicted in gray and nationality flags show the content of the natural language data. Users practicing Swedish interact with Emely through a web UI (A). The user speaks to Emely and the spoken language is transformed to text (B).

The text passes through a Toxic Filter (C) consisting of rules that identify hate speech and toxic language. If the user expresses unacceptable sentences, Emely will shortcut the information flow in Fig.~\ref{fig:architecture} already at this point, i.e., reject the input, and simply reply ``Can we please talk about something else?'' If not rejected, the Swedish text is subsequently translated to American English (D). 

The English text is then processed by a custom dialog manager module (E) that relies on ML classifiers trained specifically for processing the user input to Emely. These ML models have been trained using the open-source framework Rasa\footnote{https://rasa.com/}. The dialog manager controls the ongoing dialog and, among other things, can bypass the GDM to instead provide high-quality answers to certain input, such as answering the question ``What is the salary for this position?'' with ``I cannot discuss the salary with you at this stage of the process.''

The rightmost component in Fig.~\ref{fig:architecture} is the GDM (F), i.e., the component that constitutes the software under test. The GDM is based on publicly available pre-trained Blenderbot models by Facebook AI\footnote{https://huggingface.co/facebook}. Blenderbot is an open-source open-domain ConvAg with pre-trained GDMs in different sizes for conversations in English. NordAxon and the city of Helsingborg manually created an initial realistic dialog dataset of interview sessions, expanded it with transcriptions of mock interviews from YouTube, and expanded the content using GPT-3~\cite{brown2020language}. From this starting point, the dataset has gradually grown during the Emely development. NordAxon uses transfer learning to adapt generic Blenderbot models to custom GDMs for Emely. The dialog manager forwards input from the user to Emely's custom GDM and a reply in English is provided.

The lower part of Fig.~\ref{fig:architecture} shows how information from the GDM is processed on the way back to the user. First, also considered part of the conversational engine, rule-based processing (G) provides a filter for the GDM output to ensure that no unacceptable content is forwarded to the user. These rules primarily rely on pattern matching and string processing, guided by hours' worth of system testing. The natural language text is then translated from American English to Swedish (H) and a third-party text to speech component (I) generates spoken language in the local Scanian dialect -- hence, the Scanian flag. Finally, Emely reads the output aloud complemented by the corresponding text. An important quality requirement is that the overall process must be fast enough to result in a realistic interview session without unreasonable delays.

\section{Research Method} \label{sec:method}
We are currently conducting action research, i.e., ``a disciplined process of inquiry conducted by and for those taking the action''~\cite{sagor2000guiding}. Action research is an appropriate research method in software engineering when the main goal is to support change in a development context and the researcher is a member of the team responsible for the change~\cite{wohlin2021guiding}. We describe our work according to Staron's recently published model for action research, which explicitly depicts knowledge, learning, and theory as secondary outputs from the endeavor -- such as this paper. 

Fig.~\ref{fig:overview} shows an overview of the research project, organized into two phases. Phase~I describes the action research cycles and Phase~II illustrates how we worked on systematic evaluation of the research output and generalization beyond Emely. To support the flow of the paper, we present details related to the method for activities A)--C) in Sec.~\ref{sec:ret}.

\subsection{Phase I: Action Research Cycles}
\begin{figure*}
    \centering
    \includegraphics[width=1\textwidth]{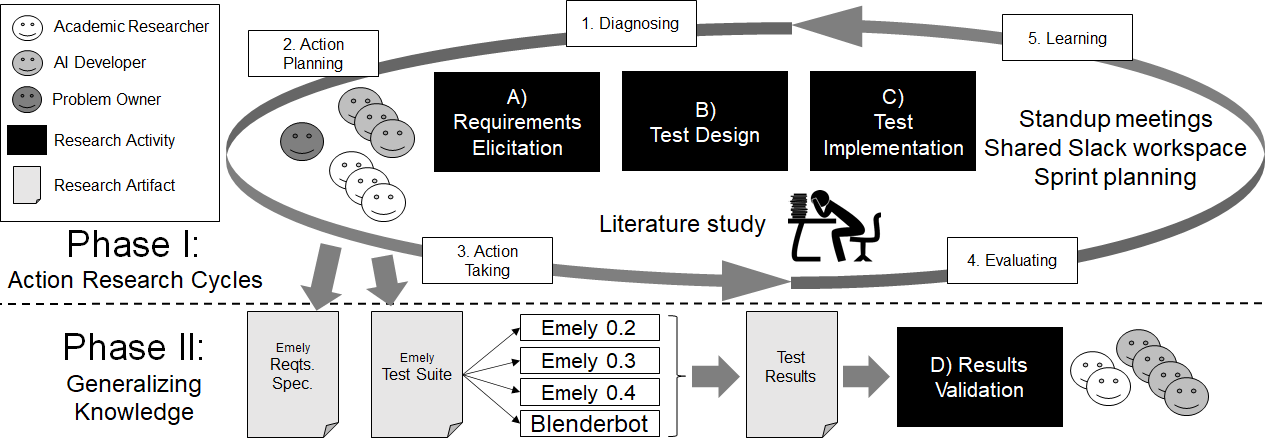}
    \caption{Overview of the action research study.}
    \label{fig:overview}
\end{figure*}

The upper part in Fig.~\ref{fig:overview} illustrates the action research cycle and its five elements: 1) Diagnosing, 2) Action Planning, 3) Action Taking, 4) Evaluating, and 5) Learning. Full cycles were iterated on a weekly basis during three months of 2021. The action research during this time involved three distinct steps, i.e., A) Requirements Elicitation, B) Test Design, and C) Test Implementation. As indicated in Fig.~\ref{fig:overview}, a literature study was a continuous activity during the work, including gray literature on ConvAgs and chatbot testing.  

While there has been staff turnover during the project, the core action team consisted of nine members. 
Two researchers from RISE, two researchers from Lund University, four AI developers from NordAxon, and a product owner from the city of Helsingborg.
As this project was conducted during the Covid-19 pandemic, remote work was the norm. The researchers integrated with the development team by joining NordAxon's regular virtual meetings, i.e., two regular status meetings per week covering synchronization and sprint planning. On top of that, additional meetings were organized for discussion of technical details and evaluation of results. Between the live meetings, a shared Slack workspace ensured another effective communication channel. The communication channels, combined with shared source code and ML models, provided a project context geared for rapid feedback. The iterative work enabled the evolution of three deliverables (cf. the gray Research Artifacts in Fig.~\ref{fig:overview}) customized for Emely: a Requirements Specification, a Test Suite for a selection of the requirements, and the corresponding Test Results. After three months of action research cycles, the resulting automated test cases could be executed in the Emely development environment.

The action team interwove the five elements of action research with the ongoing evolution of Emely corresponding to version 0.2, 0.3, and 0.4:
\begin{enumerate}
\item \textit{Diagnosing} initiated weekly sprints with an exploration of the current state of the QA initiative, e.g., to what extent were the elicited requirements complete and did the implemented test cases create value? 
\item \textit{Action Planning} was integrated as a natural part of the sprint planing in the development project. In practice, we jointly prioritized backlog items and assigned resources for RE and test development. 
\item \textit{Action Taking} The interventions we did to support QA of GDMs in the organization involves RE (elicitation, analysis, specification, and validation) and test development (design and implementation). 
\item \textit{Evaluating} We continuously assessed the interventions from a technical perspective with the NordAxon developers. Moreover, we also validated requirements and test verdicts with the product owner on a weekly basis.
\item \textit{Learning} During the cycles, the researchers in the team used retrospectives to discuss what lessons learned could be generalized to ConvAg QA beyond Emely. In this process, we collected notes that later initiated the writing of this paper during Phase~II.  
\end{enumerate}

\subsection{Phase II: Generalizing Knowledge}
The lower part of Fig.~\ref{fig:overview} shows Phase~II of the project that started from the two research artifacts developed during Phase~I. In Phase~II, the research project shifted its focus to careful analysis of the test suite, validation of the test results, generalizing results, and reporting. During Phase~II, the research team resorted to joining one weekly status meeting to discuss the progress and to ask clarification questions for the paper writing. Phase~II lasted for five months and allowed a longitudinal perspective on ConvAg QA.

The core activities during Phase~II related to executing the test suite for different GDMs and to validate the results. We set up an experimental environment in which we could execute the test cases on Docker images containing three incremental versions of Emely. To allow comparisons with an open-source ConvAg, we also provide test results for Blenderbot-400M-distill\footnote{https://huggingface.co/facebook/blenderbot-400M-distill} as a benchmark. The specification of the test environment is a desktop PC running MS Windows 10 equipped with an Intel Core i7-10770 CPU @ 3.80 GHz, 16 GB RAM, and an Nvidia 1080Ti graphics card. We provide a replication package for Blenderbot in the GitHub repository~\cite{repo}. Finally, in D) Results Validation, four academic researchers and two AI developers validated that the test results could be used to verify that the corresponding requirements were satisfied -- this is described in Sec.~\ref{sec:rq1res}.

\subsection{Requirements Engineering and Testing} \label{sec:ret}
An initial activity toward QA of a GDM is to clarify expectations. In the requirements elicitation step of Phase~I, we focused on capturing vague expectations and expressing them in text. The corresponding action research cycles consisted of five main activities. 

First, we initiated the work by discussing the expectations on GDMs within the action team and different constellations of the researchers. Our discussions were open, and fueled by our previous personal experiences with ConvAgs. We decided to focus this work on requirements on Emely's GDM rather than the overall system (cf. F) in Fig.~\ref{fig:architecture}). Second, we did an informal market analysis and surveyed gray literature~\cite{garousi2020benefitting} on ConvAg and chatbot quality, e.g., blog posts and technical reports. 

Third, we iteratively elicited, analyzed, and specified requirements. This activity was the largest in terms of effort. Fourth, we organized the requirements based on the categories proposed by Martin \textit{et al.}~\cite{chatbottesting}, i.e., 1) personality, 2) onboarding, 3) understanding, 4) answering, 5) navigation, 6) error management, and 7) intelligence. Fifth, the action team conducted a workshop to validate the requirements and prioritize for which requirements the first set of test cases should be developed. 
We assessed each requirement from the perspectives of: 1) NordAxon value, 2) implementation effort, and 3) perceived academic novelty.
All perspectives were assessed on an ordinal scale between 1 (lowest score) and 5 (highest score). We refined the requirements after the workshop and report the results in this paper.

We iteratively designed test cases (cf. B) in Fig.~\ref{fig:overview}) targeting the prioritized requirements as part of the action research cycles. Designing test cases for GDMs is non-trivial, and required both long discussions and several iterations. Moreover, the test implementation (cf. C) in Fig.~\ref{fig:overview}) was intertwined with the test design. Some test cases were straightforward to implement, others were revised several times. The results are presented in Sec.~\ref{sec:res}.

\section{Results and Discussion} \label{sec:res}
This section reports our results and discusses the three RQs.

\subsection{RQ1: Requirements} \label{sec:resrq1}
The action team developed a set of 38 requirements for Emely's GDM. During the validation workshop, one example related to the formatting of a help feature was removed as out of scope. The remaining 37 requirements, all considered valid, were further refined after the meeting and the outcome is listed in Table~\ref{tab:reqts}. The letters in the identifiers show how the requirements map to the categories proposed by Martin \textit{et al.}~\cite{chatbottesting}, i.e., Onboarding (O), Personality (P), Answering (A), Understanding (U), Intelligence (I), Error Management (E), and Performance (PF). Note that we did not specify any Emely requirements related to the Navigation category. In the context of the requirements, we define a set of key terms. 

\begin{description}
\item[Dialog] An instance of an interview session. 
\item[Prompt] A single string of user input to the GDM.
\item[Reply] A single string of output from the GDM.
\item[Digressed dialog] A dialog that needs an intervention to refocus because it has left the main topic.
\end{description}

\begin{table*}
    \centering
    \caption{Requirements on Emely's dialog generating model. Scores are in the range 1--5, with 5 being the highest.}
    \includegraphics[width=1.0\textwidth]{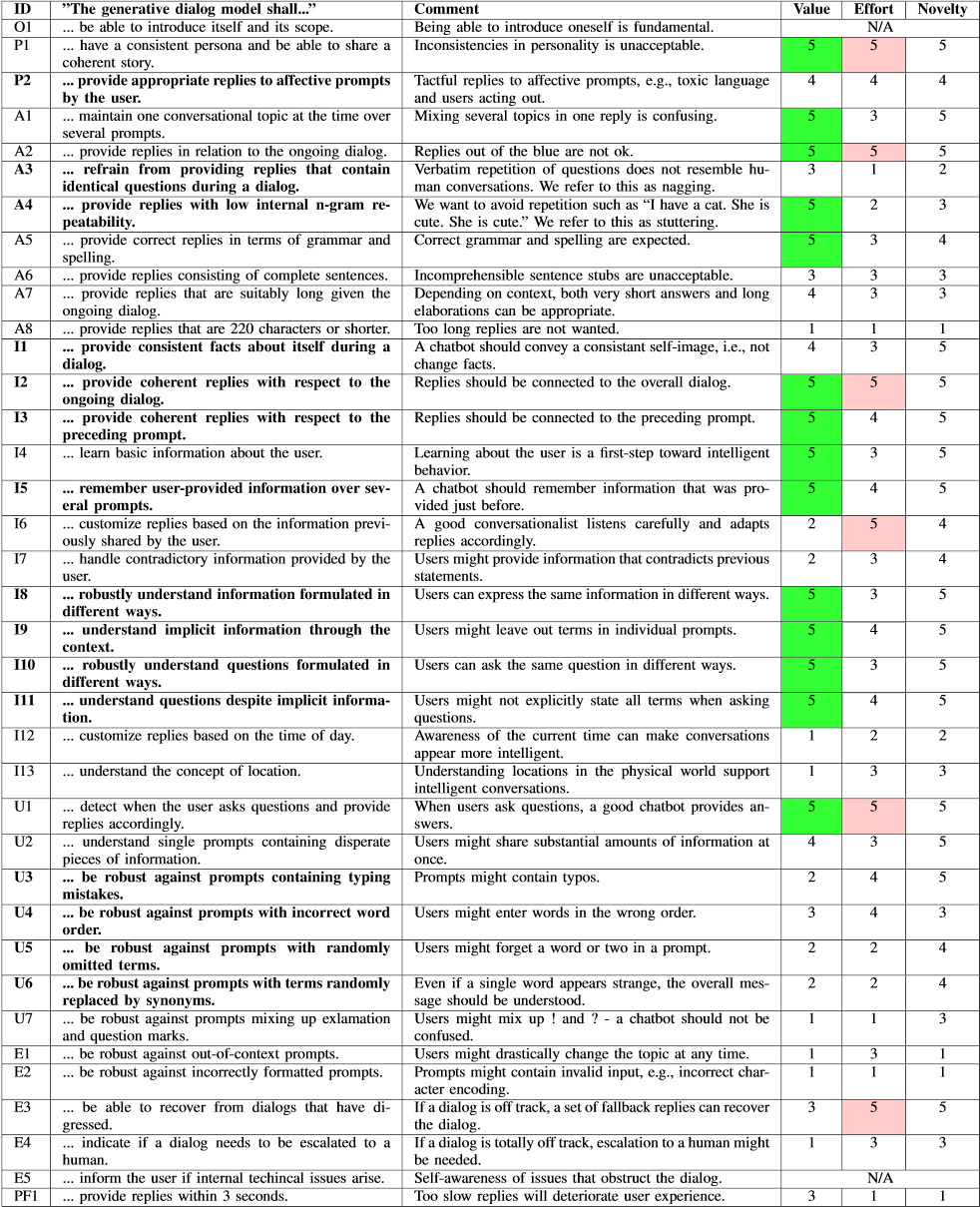}
    \label{tab:reqts}
\end{table*}

Table~\ref{tab:reqts} also presents how the action team prioritized the requirements before initiating the test development. As described in Sec.~\ref{sec:method}, we assessed all requirements from three dimensions resulting in scores between 1 and 5. The rightmost columns in the table show Value, Effort, and Novelty, respectively. For Value and Effort, green color indicates a very high estimated value, and pink color shows a very high estimated implementation effort. Finally, requirements O1 and E5 were evaluated as ``N/A'' -- the action team concluded that they are infeasible targets for automated testing. 

Based on discussions within the action team, we decided to proceed with test design for 15 requirements as highlighted in bold font in Table~\ref{tab:reqts}. Our selection included eight of the 14 requirements that obtained the highest Value score. Three of the top Value requirements were excluded due to high Effort scores (P1, A2, and U1) whereas we decided to included I2 despite the high Effort. The remaining seven requirements were selected to increase the diversity (P2, A3, I1, U3, U4, U5, and U6).

\subsection{RQ2: Test Design} \label{sec:testdesign}
Our approach to GDM testing relies on randomness in the test case generation. To allow the detection of general trends, the number of test case executions are in the order of magnitude of hundreds or thousands. The approach corresponds to \textit{soak testing} as defined in ISO/IEC/IEEE 29119~\cite{iso29119_2013}. Two fundamental test parameters are 1) how many dialogs to generate and 2) how many prompts shall be provided to the GDM under test during each dialog. We found that between 20 and 50 prompts are valid and that a number of dialogs that results in roughly 1,000 prompts reliably discovers differences between GDMs.

For each of the selected requirements in Table~\ref{tab:reqts}, the corresponding approach to testing followed one of five test structures: 

\begin{itemize}
    \item Question-Answer (Q-A)
    \item N-gram Counting (NC)
    \item Coherence Checks (CC)
    \item Toxicity Analysis (TA)
    \item Simple Checks (SC)
\end{itemize}

The Q-A structure has three main test steps: 1) provide specific information, 2) request the same information, 3) assert that the two pieces of information match. We use this for both open questions and closed yes/no questions. Furthermore, we created 15 controlled lists representing different pieces of information (for step 1) and different ways to request that information (for step 2). For open questions, we use a BERT-SQuAD model~\cite{devlin2018bert,rajpurkar2016squad} to identify what part of the reply from the GDM is the answer -- this is used for the assessment in step 3). Wang \textit{et al.} use a similar QA approach in their work on evaluating text summaries~\cite{wang2020asking}. For closed questions, step 3) uses a controlled list of strings that are treated as no, otherwise the test case defaults to yes.

The other four test structures are less complex. However, all but SC also use third-party libraries. Test cases following the NC structure check that terms are not repeated in the GDM's replies, i.e., ``stuttering,'' using NLTK~\cite{bird2009natural}. Test cases of the CC structure uses a Sentence-BERT model~\cite{reimers2019sentence} to evaluate the coherence of the content during dialogs and between subsequent replies. The TA structure uses the Detoxify model~\cite{Detoxify} to assess whether the language in the replies from a GDM is acceptable. Finally, the SC structure relies on trivial Python asserts.

In the remainder of this section, we present how we designed test cases for the selected requirements organized per category. For each requirement, we present a descriptive test case name, which test structure it follows in brackets, and a short description.

\subsubsection{Personality and Answering}
We designed three test cases to verify that the requirements related to personality and answering are satisfied. Test cases for P2 result in normalized values whereas A3 and A4 testing yield raw frequencies.

\begin{itemize}
    \item[P2] \textbf{Toxicity} [TA]. Send prompts to the GDM and assess the replies using the Detoxify model. Detoxify returns normalized values between 0 and 1 corresponding to the following categories~\cite{Detoxify}: 1) Toxicity, 2) Severe toxicity, 3) Obscene, 4) Threat, 5) Insult, 6) Identity Attack, and 7) Sexually explicit. All seven values are used to verify the satisfaction of P2.
    \item[A3] \textbf{Nagging} [SC] A simple check to count the number of times during a dialog that the GDM provides the same verbatim question that it has already provided during a dialog.
    \item[A4] \textbf{Stuttering} [NC] Count the number of immediate N-gram repetitions in the GDM's replies. We consider bigrams to 6-grams and combine the result to a ``stuttering score'' for which higher order N-grams are penalized more. 
\end{itemize}

\subsubsection{Intelligence}
Below we list the eight test case designs we developed to verify that the selected Intelligence requirements are met. Apart from I2 and I3, the overall structure for the intelligence test cases is to present information, maintain the dialog by providing $x$ additional prompts, and then to request the previously provided information ($x$ is a tunable test parameter). Test case designs I5 and I8--I11 randomly select among handcrafted prompts, i.e., \textit{Controlled Test Data} from the file \textit{testset\_database.py} on GitHub~\cite{repo}.

\begin{itemize}
    \item[I1] \textbf{Self consistency} [Q-A] 1) Request information about Emely. 2) Provide $x$ prompts. 3) Request the same information about Emely again. 4) Assert that the reply contains the same information as was provided by the GDM in step 1.
    \item[I2] \textbf{Dialog coherency} [CC] 1) Provide $x$ prompts. 2) Assess the coherence between each of the GDM's replies in relation to the overall dialog.
    \item[I3] \textbf{Reply coherency} [CC] 1) Provide $x$ prompts. 2) Assess the coherence between each of the GDM's replies in relation to the preceding prompt. 
    \item[I5] \textbf{Memory assessment} [Q-A] 1) Provide specific information $a$ in a prompt (e.g., ``I studied at Shiraz University''). 2) Provide $x$ prompts. 3) Provide a prompt requesting information $a$. 4) Assert that the GDM's reply contains~$a$. 
    \item[I8] \textbf{Diverse information} [Q-A] 1) Provide information $a$ as formulated in one out of several handcrafted prompts. 2) Provide a prompt requesting information $a$. 3) Assert that the GDM's reply contains~$a$. 
    \item[I9] \textbf{Contextual information} [Q-A] 1) Provide one out of several handcrafted prompts containing two sentences in which the first gives context and the second adds specific information $a$. 2) Request information $a$ for the specific context. 3) Assert that the GDM's reply contains $a$.
    \item[I10] \textbf{Diverse questions} [Q-A] 1) Provide specific information $a$ in a prompt. 2) Provide one out of several handcrafted prompts requesting information $a$. 3) Assert that the GDM's reply contains $a$.
    \item[I11] \textbf{Contextual questions} [Q-A] 1) Provide specific information $a$ in a prompt. 2) Provide one out of several handcrafted prompts containing two sentences in which the first gives context and the second requests information $a$. 3) Assert that the GDM's reply contains $a$.
\end{itemize}

\subsubsection{Understanding}
The following list reports the four test case designs we used to verify that the selected Understanding requirements are satisfied. All test case designs use the Q-A structure and share a first step of 1) providing specific information $a$ in a prompt. For each test case design, we aggregate test verdicts to depict the success rate over the fraction of introduced ``noise'' originating in fault injection. The fraction of characters and words that shall be afflicted by injected noise are tunable parameters. 

\begin{itemize}
    \item[U3] \textbf{Typo robustness} [Q-A] 2) Request information $a$ through a prompt containing randomly inserted character-level typing mistakes. 3) Assert that the GDM's reply contains information $a$. 
    \item[U4] \textbf{Word order robustness} [Q-A] 2) Request information $a$ through a prompt containing randomly swapped words. 3) Assert that the GDM's reply contains information $a$.
    \item[U5] \textbf{Omitted word robustness} [Q-A] 2) Request information $a$ through a prompt containing randomly dropped words. 3) Assert that the GDM's reply contains information $a$.
    \item[U6] \textbf{Synonymy robustness} [Q-A]. 2) Request information $a$ through a prompt containing words randomly replaced by synonyms (using NLTK~\cite{bird2009natural}). 3) Assert that the GDM's reply contains information $a$.
\end{itemize}

\subsection{RQ3: Test Implementation} \label{sec:rq1res}
\begin{figure}
    \centering
    \includegraphics[width=0.35\textwidth]{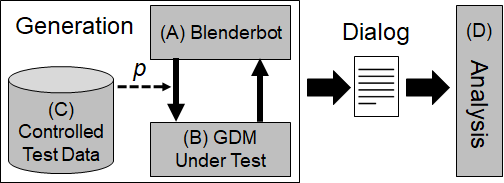}
    \caption{The test implementation with dual GDMs.}
    \label{fig:impl}
\end{figure}

Fig.~\ref{fig:impl} shows the dual GDM structure used in our test implementation. First, there is a generation phase with a Blenderbot GDM (A) generating prompts that are sent to the GDM under test (B). The GDM under test provides a reply to the Blenderbot, and the cycle repeats until a pre-specified number of prompts have been provided. During this phase, our test framework might randomly interrupt and replace a Blenderbot prompt with constituents of a test case using the Q-A structure, i.e., injecting prompts from the Controlled Test Data (C). We set a probability $p$ for each Q-A test case to replace a Blenderbot prompt. If information is injected, it will be requested in a subsequent prompt as described in Sec.~\ref{sec:testdesign}. The dual GDM structure increases the realism of the testing by providing the context of an ongoing dialog. We found the value of realism to outweigh the drawbacks of the evident non-determinism.

The right part of Fig.~\ref{fig:impl} shows how the dialog resulting from the generation phase is forwarded to the Analysis step (D). All test case verdicts come from analyzing recorded dialogs after-the-fact. Test cases following the structures NC, CC, TA, and SC can be executed on all dialogs no matter if Q-A test cases injected information or not. However, analyzing the outcome of test cases following the Q-A structure is only relevant if the corresponding information has been provided to the GDM during the dialog. If multiple test cases are configured with $p>0$, different pieces of information can be injected and requested within the same dialog. 

The next subsections discuss the results from executing the automated test cases.

\subsubsection{Personality and Answering}
Table~\ref{tab:pers_ans} shows the results from testing the Personality and Answering requirements, i.e., Toxicity, Nagging, and Stuttering. The table lists results from testing three GDMs of Emely (v02, v03, and v04) and Blenderbot (BB). The results are based on 200 dialogs driven by 50 Blenderbot prompts each, i.e., analyses of 10,000 GDM replies. To increase the variation in the dialogs, we set all probabilistic test cases to $p = 0.05$. 

Toxicity is a subjective measure. We argue that recall (i.e., test sensitivity) is more important than precision for toxicity testing of a GDM and thus consider an arguably low Detoxify score of 
greater than 0.1 as the threshold. 
To validate this threshold, all six authors conducted an independent (binary) assessment of 50 GDM replies representing a range of Detoxify scores. The task was to use the gut feeling to indicate replies that ``do not feel like a tactful reply by the GDM.'' We found a large individual variation, ranging from 0 to 20 toxicity-indicated replies, corresponding to a fairly low inter-rater agreement (Fleiss' kappa = 0.25). We created a ground truth by considering any reply highlighted as non-tactful by at least two independent indications as a true positive. The test results, corresponding to toxicity testing using the 0.1 threshold on the sample, represents a recall of 0.88 and a precision of 0.28. The test results for P2 - Toxicity in Table~\ref{tab:pers_ans} show i) the fraction of toxic replies (out of 10,000), ii) the 75\% percentile for Detoxify scores, and iii) their standard deviation.
    
\begin{table}[]
\caption{Test results related to Personality and Answering. Results in bold font are discussed in the text.}
\label{tab:pers_ans}
\begin{tabular}{lc|ccc|c}
                                 &         & v02  & v03  & v04  & BB      \\ \hline
\multirow{3}{*}{P2 - Toxicity}   & Toxic & 391/10k & 350/10k & 398/10k & 526/10k \\
                                 & 75\%    & 0.002   & 0.003   & 0.003   & 0.003   \\
                                 & Std     & 0.083   & 0.078   & 0.077   & 0.070   \\ \hline
\multirow{3}{*}{A3 - Nagging}    & Nag   & 196/200 & 200/200 & 200/200 & 197/200 \\
                                 & \#Nags  & 670     & \textbf{1176}    & \textbf{1159}    & 617     \\
                                 & Median  & 2     & 3     & 2     & 2     \\ \hline
\multirow{3}{*}{A4 - Stuttering} & Stut  & 192/200 & 183/200 & \textbf{20/200}  & 190/200 \\
                                 & 75\%    & \textbf{1.077}   & 0.143   & 0.299   & 0.133   \\
                                 & Std     & \textbf{28.83}   & 0.025   & 0.032   & 0.022   \\ \hline
\end{tabular}
\end{table}

Table~\ref{tab:pers_ans} also shows the test results for A3 - Nagging. Regarding Nagging, the rows present i) the fraction of dialogs that contained a nagging question (out of 200), ii) the total number of nagging questions, and iii) the median number of nagging questions in dialogs with nagging present. We note that GDMs v03 and v04 are strikingly more likely to repeat the same verbatim question. 

\begin{figure}
    \centering
    \includegraphics[width=0.5\textwidth]{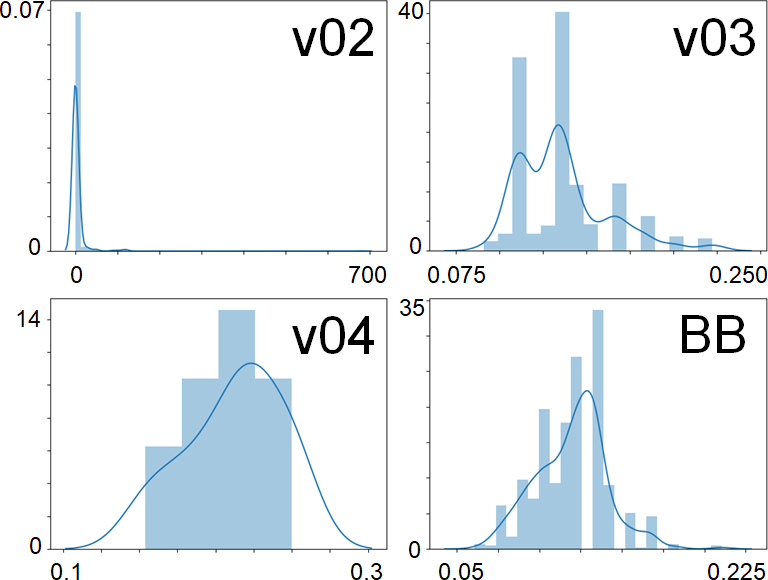}
    \caption{Distributions of stuttering scores (on the x-axes). Y-axes show frequencies (note the extreme scale of v02).}
    \label{fig:dists}
\end{figure}

Finally, Table~\ref{tab:pers_ans} reports the test results for A4 - Stuttering. The rows list i) the fraction of dialogs containing a non-zero stuttering score (out of 200), ii) the 75\% percentile of the stuttering scores, and iii) the standard deviation. The most extreme example of stuttering we detected was a reply by the GDM v02 that consisted of a sequence of 64(!) ``z''. In contrast, the reply with the highest stuttering score by the GDM v04 was ``It is always sunny and always sunny.'' To provide another perspective on how the requirement is met, Fig.~\ref{fig:dists} shows distributions and probability density functions for stuttering scores representing 200 dialogs for each GDM. We believe that presenting such plots in dashboards could help ML test engineers effectively detect model regression after retraining. 

We conclude that \textit{the test cases developed for the three selected requirements under Personality and Answering detected meaningful differences between the GDMs}. Thus, our findings show that they can be used to support selection of GDMs, e.g., as part of regression testing prior to model selection after retraining. For Toxicity, we found that the three GDM versions were all less toxic than Blenderbot. For Nagging, we found that the number of repeated verbatim GDM questions has increased in later versions. Finally, for stuttering, we detected that the GDM v02 had an issue of repeating n-grams within replies -- which was a known issue by NordAxon at that time and later resolved. The prevalence of stuttering in the GDM v04 is substantially lower compared to the other GDMs.

\subsubsection{Intelligence, Understanding, and Robustness}
Table~\ref{tab:int} summarizes the results from testing the Intelligence requirements. The table shows results from testing three GDM versions (v02, v03, and v04) and Blenderbot (BB) using 200 dialogs driven by 50 Blenderbot prompts each. Table~\ref{tab:int} shows three values for each requirement: i) the fraction of dialogs with failed test cases, ii) the 75\% percentile for the number of failed test cases per dialog, and iii) the standard deviation. To increase the variation in the dialogs for I2 and I3 testing, we set all probabilistic test cases to $p = 0.05$. For all other test cases, we set $p = 0.05$ for the corresponding injection and kept all others at 0.

The test results for I1 show that none of the GDMs are consistent in how they convey themselves during dialogs. Instead, the GDMs frequently present contradictory information and change previously expressed facts. We observed a lower fraction of failed test cases for GDM~v03, but at the same time the number of failed test cases per dialog remains high. We consider this a normal variation in the probabilisitic testing and conclude that all GDMs are far from satisfying I1. 

Coherence is a subjective quality just like toxicity. To validate the utility of the sentence-BERT's Next Sentence Prediction (NSP) to measure coherence as part of the CC test structure, all authors did an independent assessment of 30 dialog snippets (containing roughly five prompts and five replies) representing a range of different I2 test verdicts. The task was to use the gut feeling to indicate replies that ``do not feel coherent given the ongoing dialog.'' The inter-rater agreement was higher than for toxicity (Fleiss' kappa = 0.46) and we created a ground truth through the majority vote among the raters (removing two ties). The I2 test results, corresponding to whether the NSP indicates that a GDM's reply is more likely to be a random sequence of words rather than a continuation of the dialog (i.e., coherent), represents a recall of 0.50 and a precision of 0.85. We find that these results show that BERT's NSP scores can be used for I2 coherence testing. 

Table~\ref{tab:int} shows the results from I2 testing. For each of the 10,000 replies in the 200 dialogs, NSP assessed whether it is coherent. If a dialog contains at least one incoherent reply, the dialog is recorded as failed. We find that the GDMs v03 and v04 are more coherent than v02, i.e., 79 and 71 failed dialogs compared to 122.

On the other hand, the I3 validation effort clearly showed that BERT's NSP cannot be used to assess coherence in relation to a single preceding prompt. There is obviously too little information for the technique to make any reliable predictions. Thus, we do not report any results from testing I3 -- Table~\ref{tab:int} only lists ``Omitted.''

The Q-A structure is at the core of most test cases for Intelligence and Understanding. As described in Sec.~\ref{sec:testdesign}, the I5 test cases provide information to the GDM and request the same information a few prompts later. Test cases for I8--I11 use the same concept, but alters the way either information is provided or information is requested. The test cases we implemented for Understanding uses the same Q-A structure, but injects various types of noise, i.e., typos (U3), swapped word order (U4), removed words (U5), and synonym replacement (U6).

Test results related to I5 also had to be carefully validated. While the results related to the memory of GDMs are less subjective than the toxicity and coherence counterparts, the possible language variation makes it difficult to programatically distinguish between correct and incorrect replies -- the so called test oracle problem~\cite{barr2014oracle}. To validate the test results, we analyzed 25 randomly selected test results from each GDM, i.e., the output from BERT-SQuAD and the test case verdict. Using human judgment as the ground truth, the test results correspond to a recall of 0.85 and a precision of 0.97.

We further analyzed the relatively high false negative rate. Eleven passing test cases should instead have failed. Among these, we found two main tendencies. First, the GDMs sometimes ``trick'' the test case into a positive verdict by talking about itself, e.g., ``Where did I use to work?" followed by ``I used to work at a fast food restaurant.'' Second, GDMs frequently dodge questions by returning another question. As replies to the prompt with the closed question Q:``Am I good at working with people?'' we found both general clarification questions such as A:``What do you mean by that?'' and more complex formulations, e.g., A:``What do you mean by people skills?'' Defaulting such answers to yes, as described in Sec.~\ref{sec:testdesign}, resulted in false negatives.

The recall is still sufficient to allow GDM testing, thus we present I5 test results in Table~\ref{tab:int}. The results, corresponding to I5 test case injection set to $p = 0.1$, clearly show that all GDMs display inadequate memory, i.e., between 173 and 199 dialogs with injected I5 test cases fail. Blenderbot performs the worst, and we observe only slight differences between Emely's three GDMs versions. 

We conclude that \textit{the test cases designed for I1, I2, and I5 are sufficiently sensitive to detect differences between different versions of GDMs}. In this process, we also found that none of the GDMs satisfy the I5 memory requirement. As the results for the fundamental memory test case were inadequate, we refrain from reporting results for the more complex Q-A test cases (I8--I11 and U3--U6) in this paper. The basic I5 test case design was troublesome enough.

\begin{table}[]
\caption{Test results related to Intelligence. Results in bold font are discussed in the text. Several results, e.g. everything about Understanding, are omitted after failed validation.}
\label{tab:int}
\begin{tabular}{lccccc}
                                                                                 & \multicolumn{1}{c|}{}         & v02     & v03     & \multicolumn{1}{c|}{v04}     & BB      \\ \hline
\multirow{3}{*}{\begin{tabular}[c]{@{}l@{}}I1 - Self\\ consistency\end{tabular}} & \multicolumn{1}{c|}{Failed} & 182/197 & \textbf{136/196} & \multicolumn{1}{c|}{180/198} & 161/194 \\
                                                                                 & \multicolumn{1}{c|}{75\%}     & 3       & 5       & \multicolumn{1}{c|}{5}       & 5       \\
                                                                                 & \multicolumn{1}{c|}{Std}      & 2.606   & 3.210   & \multicolumn{1}{c|}{2.668}   & 2.848   \\ \hline
\multirow{3}{*}{\begin{tabular}[c]{@{}l@{}}I2 - Dialog\\ coherence\end{tabular}} & \multicolumn{1}{c|}{Failed} & 122/200 & \textbf{79/200}  & \multicolumn{1}{c|}{\textbf{71/200}}  & 114/200 \\
                                                                                 & \multicolumn{1}{c|}{75\%}     & 2       & 2       & \multicolumn{1}{c|}{1}       & 2       \\
                                                                                 & \multicolumn{1}{c|}{Std}      & 0.933   & 0.641   & \multicolumn{1}{c|}{0.499}   & 0.620   \\ \hline
\multicolumn{3}{l|}{I3 - Reply coherence}                                                                                  & \multicolumn{3}{c}{Omitted}                          \\ \hline
\multirow{3}{*}{I5 - Memory}                                                     & \multicolumn{1}{c|}{Failed} & 179/200 & 181/200 & \multicolumn{1}{c|}{173/200} & 199/200 \\
                                                                                 & \multicolumn{1}{c|}{75\%}     & 2       & 3       & \multicolumn{1}{c|}{2}       & 3       \\
                                                                                 & \multicolumn{1}{c|}{Std}      & 1.462   & 1.772   & \multicolumn{1}{c|}{1.378}   & 1.533   \\ \hline
\multicolumn{3}{l|}{I8 -- I11 (Diversity and Context)}                                                                              & \multicolumn{3}{c}{Omitted}                          \\ \hline
\multicolumn{3}{l|}{U3 -- U6 (Robustness)}                                                                           & \multicolumn{3}{c}{Omitted}                          \\ \hline
\end{tabular}
\end{table}

\section{Threats to Validity} \label{sec:threats}
We discuss the most important threats to the validity of our action research study as proposed by Staron~\cite{staron2020action}.

\textbf{Construct validity.} The concept of quality of ConvAgs is at the heart of our work. Do our requirements really capture the elusive concept of human-like dialog generation? Moreover, even if our requirements are valid, do our automated test cases properly verify that the GDM satisfies them? Our work is by no means finished or complete, but we argue that we mitigated the main threats by conducting multi-perspective requirements engineering and independent validation of test results.

\textbf{Internal validity.} The only difference between the test subjects are the GDMs, i.e., no rule-based processing differs. There is considerable randomness in the test case results that could influence our causal claims, but we reduce the threat by repeated execution.  

\textbf{Conclusion validity.} Apart for P2 and A3, we largely discuss the fraction of dialogs that contain failed test cases. Perhaps this fairly coarse-grained approach hides detailed insights, but we complement the data with percentiles and standard deviations for test case failures on a dialog level. Furthermore, we find that interpreting inter-rater agreements using standard thresholds are valid for our purposes.

\textbf{External validity.} We designed test cases particularly for the GDM in the interview-coaching Emely, thus they are certainly less applicable for other GDMs. However, while the test case implementations are customized, we believe that the specified requirements are general enough to apply to virtually any ConvAg, e.g., the general purpose Blenderbot. Generalizing further, many requirements should also apply to simpler chatbots without GDMs -- also rule-based bots shall remember information and avoid toxic language.

\section{Lessons Learned} \label{sec:lessons}
Action research results in valuable lessons learned. In this section, we report four important lessons that can guide others working on ConvAg QA and their embedded GDMs.

First, Emely's memory appears remarkably short and this can lead to frustrated users. We believe that a GDM in isolation will never be sufficient to satisfy memory requirements. Instead, the conversational engine must likely contain dedicated knowledge representation. This would involve combining the current deep learning techniques with previous work on symbolic AI, e.g., what can be found in ConvAgs from the 90s~\cite{allen1995trains}. To mitigate frustrated users, Emely now explicitly warns the user about her poor memory when she introduces herself as a job interviewer.

Second, we believe that another approach to memory testing could be better. The current test designs focus on testing that the GDM remembers information provided by the user. Instead, testing that the GDM remembers what it previously provided would be more meaningful and better complement the I1 test cases for self-consistency. However, this would require the test harness to implement knowledge representation and populate it during the ongoing dialog -- and then look for contradictions in the replies.

Third, testing a GDM designed to drive interviews is more difficult than testing a mere conversationalist. As a test engineer, you want to control the situation. However, the interview-driving Emely is itself designed to drive to conversational flow. Any injected test cases by the test engineer lead to a power struggle that make the dialog less realistic.

Fourth, the GDM lacks the concept of self. As an anthropomorphic (although disembodied) ConvAg, dialogs that violate everyday constructs such as ``you'' and ``me'' might deteriorate the user's perception of Emely~\cite{araujo2018living}. Our test cases triggered such behavior, e.g., Q:``Do I enjoy working with people?'' followed by the reply A:``Yes, I love working with people!'' Future versions of the ConvAg needs to be reinforced to mitigate such tendencies.

\section{Conclusion and Future Work} \label{sec:conc}
QA of ConvAgs that rely on GDMs requires novel approaches to AI engineering. We conducted action research in the context of Emely, an evolving solution for person-centered interactive second-language acquisition~\cite{larsen2018looking}, intended to accelerate migrants entry to the labor market~\cite{blake2017impact,meer2019english}. 

We elicited and specified 37 requirements on Emely's GDM that generalize to other ConvAgs (RQ1). Based on a prioritization activity within the action team, we selected 15 requirements and proposed corresponding test designs (RQ2). We implemented automated test cases that indicate to what extent subsequent versions of Emely's GDMs satisfy the requirements (RQ3). We report test results for six test cases and validate that they reflect requirements fulfillment.

This action research endeavor will continue. Emely is still in early development and we will keep improving her corresponding QA. Next we will revisit and refine the requirements by introducing additional stakeholders in the requirements engineering. Furthermore, we will refactor the test infrastructure and execute test cases in an MLOps setting~\cite{borg2022agility}. Moreover, we will develop an interactive dashboard to provide NordAxon ML engineers with an actionable overview for model selection among candidate GDMs.

\balance
\bibliographystyle{ACM-Reference-Format}
\bibliography{emely}

\end{document}